# Conversational Collective Intelligence (CCI) using Hyperchat AI in a Real-world Forecasting Task

Hans Schumann
Unanimous AI
San Francisco, CA, USA
Hans@Unanimous.ai

Louis Rosenberg
Unanimous AI
Pismo Beach, CA, USA
ORCID: 0000-0003-3457-1429

Ganesh Mani
Carnegie Mellon University
Pittsburgh PA, USA
ORCID: 0000-0002-2170-7414

Gregg Willcox
Unanimous AI
Seattle, WA, USA
Gregg@Unanimous.ai

*Abstract*— Hyperchat AI is a novel agentic technology that enables thoughtful conversations among networked human groups of potentially unlimited size. It allows large teams to discuss complex issues, brainstorm ideas, surface risks, assess alternatives and efficiently converge on optimized solutions that amplify the group's Collective Intelligence (CI). A formal study was conducted to quantify the forecasting accuracy of human groups using Hyperchat AI to conversationally predict the outcome of Major League Baseball (MLB) games. During an 8-week period, networked groups of approximately 24 sports fans were tasked with collaboratively forecasting the winners of 59 baseball games through real-time conversation facilitated by AI agents. The results showed that when debating the games using Hyperchat AI technology, the groups converged on "High Confidence" predictions that significantly outperformed Vegas betting markets. Specifically, groups were 78% accurate in their High Confidence picks, a statistically strong result vs the Vegas odds of 57% (p=0.020). Had the groups bet against the spread (ATS) on these games, they would have achieved a 46% ROI against Vegas betting markets. In addition, High Confidence forecasts that were generated through above-average conversation rates were 88% accurate, suggesting that real-time interactive deliberation is central to amplified accuracy.

*Keywords—collective intelligence, AI agents, Hyperchat AI, collective superintelligence, forecasting, agentic interfaces, sports*

## I. Introduction

Large organizations often rely on networked human teams to inform critical forecasts, decisions and assessments. In many applications, large teams are needed because they possess broader perspectives and provide a wider range of expertise, insight, situational awareness, and institutional knowledge. The field of Collective Intelligence (CI) is devoted to powering large-scale collaboration and has produced an expansive body of research showing that large human groups can significantly outperform small human groups when forecasting events and assessing outcomes [1].

Unfortunately, most CI tools for large-scale groupwise forecasting are based on the collection and aggregation of individually provided estimates rather than a thoughtful and interactive deliberation among participants. Specifically, most CI methods for large groups involve collecting simple datapoints via online polls, surveys, prediction markets, or other narrow data entry mechanisms. While such methods can moderately amplify group intelligence, prior research in the CI subfield of Swarm Intelligence (SI) has shown that real-time interactive deliberation among large groups can significantly improve the accuracy of group forecasts, estimations, and evaluations [2].

The first swarm-based CI method was developed in 2014 and employed a graphical interface to enable large, networked groups to collaboratively guide a pointer to a collectively selected solution [3]. Known as Artificial Swarm Intelligence (ASI) this methodology has been shown in a variety of studies to significantly outperform non-interactive methods like polls, surveys, and prediction markets [4,5] That said, ASI, like polls, surveys, and markets, only allow for limited content to be exchanged among participants, usually in the form of numerical estimates, ranked options, or multiple-choice selections. This reduces the scope of real-time groupwise deliberations and limits the richness of the forecasting output [6].

In 2023, a new CI method emerged called Conversational Swarm Intelligence (also referred to as *Hyperchat AI™*) that enables large human groups to hold real-time discussions at massive scale [7]. Hyperchat AI works by dividing a large group into a set of subgroups, each sized for thoughtful real-time discussion. Hyperchat AI technology then inserts a unique AI agent called a "Conversational Surrogate" into each subgroup. Each agent is designed to participate in its local discussion, extracting key insights as they emerge and relaying insights to AI agents in other subgroups. When agents receive insights, they express them conversationally in their local group. This weaves the local discussions together into a coherent global dialog. In this way, large, networked teams with hundreds of members can discuss issues, debate alternatives, surface risks, brainstorm ideas, and converge in real-time on unified solutions [8]

This approach has been shown to enable large groups to efficiently discuss issues, brainstorm ideas, and converge on solutions that outperform traditional CI methods [7-10]. For example, researchers at Unanimous AI and Carnegie Mellon University tasked groups of 35 people with answering IQ test questions together using a Hyperchat AI powered software platform called Thinkscape®. The study showed that hyperchat-connected groups could efficiently and accurately solve IQ questions through real-time conversations. The test groups using Thinkscape scored an average IQ in the 97th percentile (IQ = 128), which is gifted-level intelligence, and significantly outperformed the median individual in each group (IQ = 100) and traditional CI methods (IQ = 115) [10].

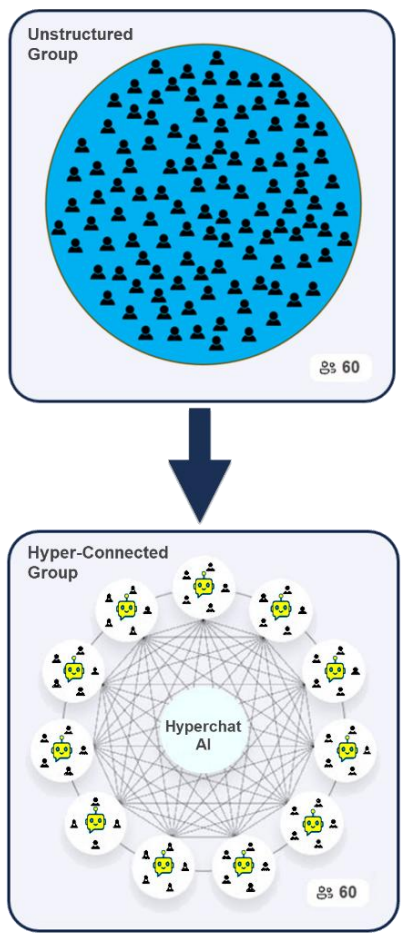

Fig. 1. Hyperchat AI turns a large group into a Conversational Network

While a variety of prior research has been conducted to measure the effectiveness of Hyperchat AI technology, no formal studies have yet been performed to quantify the value of hyperchat in groupwise forecasting. To address this need, an 8-week study was conducted in which Major League Baseball (MLB) games were collaboratively predicted by groups of sports enthusiasts using an online conversational forecasting platform powered by Hyperchat AI technology.

## II. MLB FORECASTING STUDY

During each of the **15** forecasting sessions, the outcomes of four MLB games were collaboratively predicted by groups of approximately 20 to 25 human subjects engaged in real-time conversational deliberation. In total, **60** games were collaboratively predicted during this study, but one game was rained out resulting in a final dataset of 59 forecasts. The sessions were run on Tuesday and Friday afternoons during seven consecutive weeks of the 2025 MLB season starting on July 18. The deliberations were conducted using an online collaboration platform called Thinkscape.ai that incorporates Hyperchat AI technology to allow networked human groups of up to 250 people to hold real-time conversations via standard web browsers. The Thinkscape.ai platform supports real-time discussions by text, voice, or video. For this study, the researchers used Thinkscape in text chat mode.

The human participants were sourced from a professional panel-providing service (Prolific). Each participant was paid approximately $6.50 for taking part in each session. They were selected at random for this study from a pool of candidates that self-identified as "baseball fans" to confirm they were familiar with the sport and could discuss the topic. During each forecasting session, participants were provided with a link to the Thinkscape.ai platform and were told the session would last approximately 30 minutes.

When entering the Thinkscape platform each group of 20 to 25 participants was shown a 1-minute instructional video about how the platform works. The group was then randomly divided into five subgroups (called Thinktanks), each with approximately 4 to 5 participants. As explained above, each subgroup enables real-time dialog among human members and an AI agent called a Conversational Surrogate. Each AI agent was tasked with (i) monitoring the discussion in its local subgroup, (ii) extracting insights about the game being discussed, (iii) sending insights to the AI agents in other subgroups, and (iv) expressing the insights received from agents in other subgroups to members of its own subgroup.

In this way, the network of AI agents connect the five local conversations into a single global conversation in which perspectives are debated, for or against each team, with arguments being efficiently shared across subgroups. Specifically, content is shared using an intelligent matching algorithm that gives preference to content that will challenge each subgroup with ideas, perspectives and insights that have not yet been considered among the members of that subgroup. For example, if a subgroup strongly believes that the Los Angeles Dodgers will win a game they were asked to forecast, the AI agent in that subgroup is likely to receive and express counterpoints sourced from other groups that provide arguments as to why the Dodgers may lose. This happens simultaneously in all subgroups, ensuring that key insights supporting or opposing each outcome propagate conversationally throughout the full collaborative group.

During each forecasting session, a set of four questions was asked to the group of human subjects. Each question lasted approximately 5 minutes and tasked the group with conversationally predicting which of the two teams playing would win a specific MLB game scheduled for that evening. For example, during the session run on Friday, September 12, 2025, the first question tasked the group with considering the Kansas City Royals (Michael Lorenzen pitching) game against the Philadelphia Phillies (Walker Buehler pitching).

After being presented with simple data indicating the two teams competing, the starting pitchers for each team, and listing which team was playing at their home field, the participants were asked *"Which team is most likely to win this game and by how many runs and why?"* The intent of this wording was to inspire the group to conversationally debate which of the two teams would win, how many runs the team was likely to win by, and to provide reasoning that supports each outcome argued for or against.

At the start of each question, a timer appeared on each participant's screen that counted down from 5 minutes, thereby informing them as to how much time they have left to debate the outcome of the game in question. As their conversation unfolded, each participant's dialogue was processed using a Large Language Model (LLM) to determine their current views regarding which team will win, the strength of their belief, and their reasoning. This analysis occurred in real-time for all 20 to 25 participants in the session and was continuously updated as the dialog unfolded. This allowed the Thinkscape application to track how participants shifted their perspectives (individually, and as a collective) as they were conversationally exposed to views and reasoning of others. Figure 2 below shows the change in aggregated collective support, as was captured and assessed conversationally, across the duration of a deliberation among 22 human participants using Hyperchat AI as they assessed an MLB game between the Phillies and the Rangers.

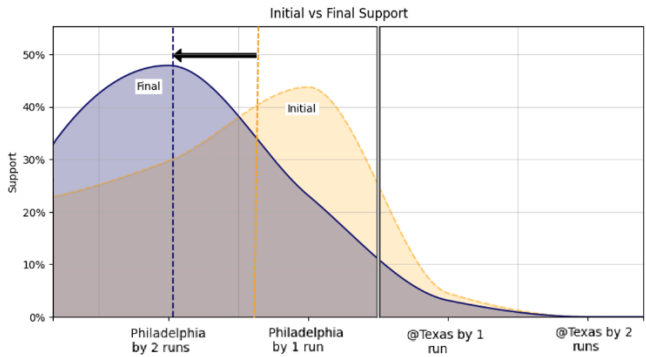

Fig. 2. Change in aggregated Conversational Support during discussion.

As each conversation unfolded, the expressed sentiments of the participants were quantified and aggregated in real-time to determine a weighted mean that reflected which team was collectively predicted by the group to win the game, and by how many runs. The weighted mean at the conclusion of the five-minute deliberation was stored as the final collective forecast. Using conventional baseball betting lines, any final collective forecast in which one team was predicted to win by 1.5 runs or more was classified as **High Confidence**, and any collective forecast in which a team was predicted to win by less than 1.5 runs was classified as **Low Confidence**. In this way, each game had four possible predicted outcomes that were reported: (i) Team A wins / Low Confidence; (ii) Team A wins / High Confidence; (iii) Team B wins / Low Confidence; and (iv) Team B wins / High Confidence

### III. RESULTS

As described above, the outcome of each of the **59** games in this study were predicted through natural conversational deliberations among networked human groups of 20 to 25 participants. Each deliberation lasted five minutes and was conducted using Hyperchat AI technology that divided the full group into subgroups interconnected by AI agents. At the conclusion of each conversational deliberation, the system output one of four possible collective predictions: Team A / Low Confidence; Team A / High Confidence; Team B / Low Confidence; and Team B / High Confidence.

After each game was played, the outcome was recorded and the forecasts were scored. In addition, the Vegas odds for each game were documented from a major sportsbook to use as predictive benchmarks. **Table I** below shows the accuracy of the forecasts generated conversationally using Hyperchat AI technology across the full set of 59 games. As shown, 46% of the games (27 games) were deemed High Confidence predictions (i.e., the group converged on a collective prediction that one team would win by 1.5 runs or more), and approximately 54% of the games (32 games) were deemed Low Confidence predictions (i.e., the group converged on a collective prediction that one team would win by less than 1.5 runs). The predictions were grouped by classification and scored for accuracy.

As shown in Table I, High Confidence predictions that were generated conversationally using Hyperchat AI technology were **78% accurate**, which is an extremely strong result across a set of 27 games that had average Vegas odds of 57%. This is a statistically significant result, as a Poisson-Binomial distribution can be used to show that the 78% accuracy across these 27 High Confidence forecasts is statistically significantly higher than the expected win percentage using the published Vegas odds (p=0.020). This is supported by confidence intervals for the true accuracy of High and Low Confidence predictions, which show that we are 95% confident High Confidence predictions are between 59.2% and 89.4% accurate and Low Confidence predictions are between 25.5% and 57.7% accurate. These intervals do not overlap, thus indicating that High Confidence predictions significantly outperform Low Confidence ones.

TABLE I    Prediction Accuracy

|  | Number of Games | Accuracy | Vegas Odds | p-value |
|---|---|---|---|---|
| Low Confidence (under 1.5 runs) | 32 | 40.6% | 53.1% |  |
| High Confidence (over 1.5 runs) | 27 | 77.8% | 57.0% | 0.020 |

To appreciate the accuracy implications of these High Confidence predictions, consider a bettor who placed a $100 wager on all 27 of these games against the Vegas odds. As shown in Table II below, this bettor would have achieved a 37% return on investment and a total profit of $997. The cumulative profit over time is shown in Figure 3 below.

TABLE II    Wager Modeling

|  | Profit | ROI | Accuracy | Vegas Odds on Bets | p-value |
|---|---|---|---|---|---|
| High Confidence Picks | $997 | 36.9% | 77.8% | 57.0% | 0.020 |
| High Confidence Picks (with ATS) | $1,245 | 46.1% | 63.0% | 44.0% | 0.037 |
| Against Low Confidence Picks | $736 | 23.0% | 59.3% | 49.1% | 0.161 |

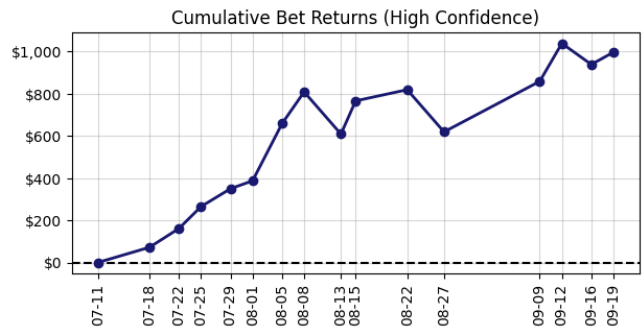

Fig. 3. Cumulative profit when wagering on High Confidence picks.

In addition, a Cohen's d analysis was also performed to assess the magnitude of the difference between High Confidence and Low Confidence predictions. The effect size was d=0.82, indicating a large practical difference between the predictions.

We can also compare performance of the High Confidence predictions to the "Against the Spread" (ATS) betting lines published by Vegas for each game. For Major League Baseball games, the typical ATS lines require that the winning team wins by 2 or more runs and therefore have a significantly higher payout. As shown in Table II above, if a bettor had placed an ATS wager on all 27 High Confidence predictions, they would have had an accuracy of 63% in those wagers and would have generated a profit of $1245 which is a 46% ROI. This betting model is also statistically significant (p=0.037). The cumulative profit over time is shown in Figure 4 below.

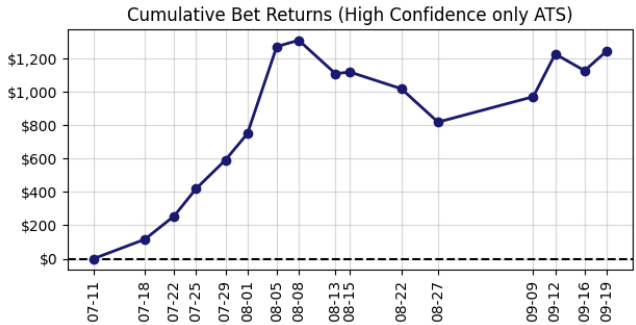

Fig. 4. Cumulative profit when wagering ATS on High Confidence picks.

Referring back to Table I and Table II, we must also consider the 32 games classified as Low Confidence predictions. These were consistently inaccurate, yielding a 41% win rate against Vegas odds that predicted a 53% win rate. This is an interesting result because it suggests that a collective forecast classified as Low Confidence can be used as an inverse signal when placing wagers. For example, had a bettor chose to bet against all of the Low Confidence picks, they would have generated a 23% ROI across the 32 games and a cumulative profit of $736. The cumulative profit over time is shown in Figure 5 below.

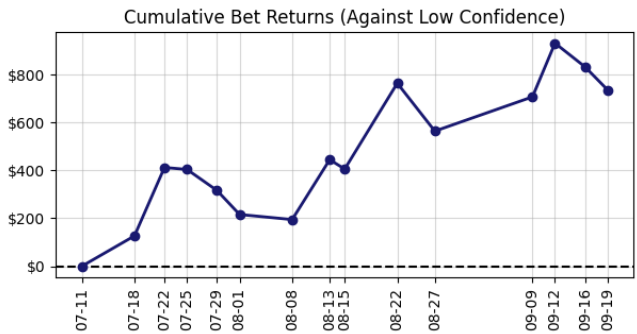

Fig. 5. Cumulative profit when wagering against Low Confidence picks.

To appreciate why the Low Confidence picks provide a strong inverted signal in which the opposite team is likely to win, we need to consider that every Vegas betting line has a "favorite" that is deemed by the oddsmakers as more likely to win. In general, data shows there is a strong (and irrational) bias among the public to bet on favorites despite the fact that the payout on underdog picks is proportionally higher [11, 12]. In this study, 43 out of the 59 teams selected to win (i.e. 73%) were the Vegas favorites. This aligns with prior research indicating a strong favorite bias among sports enthusiasts.

Combining this clear evidence of a favorite bias with the strong accuracy of High Confidence picks and the weak accuracy of Low Confidence picks we can hypothesize that the deliberative process enabled by the Hyperchat AI discussions provides a powerful signal that distinguishes between favorite picks driven mostly by deliberative reasoning (i.e. High Confidence picks) versus favorite picks driven mostly by bias (i.e. Low Confidence picks). In fact, High Confidence favorite picks (23 games) were 78% accurate in this study, while Low Confidence favorite picks (20 games) were only 41% accurate.

Clearly, the groupwise conversational forecasts generated using Hyperchat AI technology were extremely accurate when the groups converged on High Confidence picks. An important question still remains – *was this accuracy a direct result of the group's deliberations?* To explore this, we quantified the quality of each deliberation by measuring the number of messages passed among group members when considering the outcome of each game. We then compared the accuracy of conversational forecasts for games that had above-average messages per minute vs games that had below-average messages per minute. We did this for both the High Confidence predictions and the Low Confidence predictions. This resulted in the data shown in Table III below.

| TABLE III | Conversation Quality vs Accuracy | |
|---|---|---|
| | High Confidence Picks | Low Confidence Picks |
| Above Average Conversation | 16<br>87.5% | 13<br>30.8% |
| Below Average Conversation | 11<br>63.6% | 19<br>47.4% |

As shown in the table above, High Confidence forecasts that were generated through above-average conversation rates were 88% accurate compared to 64% accuracy for games that were predicted with below-average conversation rates. This suggests that when groups deliberate vigorously and converge on a High Confidence forecast, it is especially accurate compared to when the conversation is less vigorous.

Even more interesting, we measured an inverted result for Low Confidence picks. As shown in Table III, when groups deliberate with above-average conversation rates and converge on Low Confidence predictions, the forecasts are only 31% accurate. This is not as counterintuitive as it might seem. After all, these groups debated vigorously among themselves and still did not converge on a forecast with high confidence. Thus, despite sharing detailed reasoning for why one team was more likely to win versus the other, the group remained divided and unsure. Thus, it makes sense that vigorous debates resulting in low confidence forecasts would be the weakest picks. On the other hand, Low Confidence picks generated by below average deliberation (because the group didn't have that much to argue about), had a higher average accuracy of 47%. Such picks were likely "toss-up" games where there were not strong arguments for picking one team over the other.

Taken together, these results suggest that when engaging a moderately-sized networked group (20 to 25 people) in a Conversational Collective Intelligence process like that enabled by Hyperchat AI technology, the strongest forecasts are likely those in which the group converges collectively with High Confidence through a vigorous deliberation with an above-average conversation rate. In this study there were 16 games (out of 59 considered) in which High Confidence was achieved through above-average conversation. These games were predicted with 88% accuracy against 57% Vegas odds, which is a statistically significant result ($p = 0.010$).

Similarly, we find the least accurate predictions are those that converge with Low Confidence despite recording an above-average conversation rate. In this study there were 13 games (of 59) in which Low Confidence forecasts were produced through above-average conversation. These games were predicted with 31% accuracy against 53% Vegas odds. Betting against these weak picks would have achieved 69% accuracy ($p = 0.116$).

## IV. CONCLUSIONS

Conversational Collective Intelligence (CCI) generated using large-scale deliberations (as enabled by Hyperchat AI™) is a potentially powerful alternative to traditional CI methods such as polls, surveys, and prediction markets. In traditional methods, participants provide individual datapoints as numerical values, multiple-choice selections, or entering trades into a market. These datapoints are aggregated, which creates a groupwise result, but the process is not deliberative. In other words – there is not an interactive discussion and debate through which participants share arguments and reasoning in favor or against the various outcomes. Instead, participants are treated as a source of data for aggregation.

When using Hyperchat AI technology, on the other hand, rationales are shared for and against each option. These arguments are expressed conversationally by participants in local subgroups and are spread conversationally between subgroups by AI agents. Each point raised has the potential to sway some participants or inspire counterpoints from other participants. The most impactful arguments are propagated fastest by the AI agents, while arguments that fall flat, propagate slower. This results in an efficient deliberative network in which large groups can quickly converge on collective perspectives, revealing not just a forecast outcome but a level of confidence.

Furthermore, unlike traditional CI methods for engaging large groups, confidence using Hyperchat AI is <u>not</u> measured based merely on the numerical data provided by participants, but is also based on the strength of sentiments expressed through conversational deliberation as the group debates the options. The system assesses what each user says to support their views, how each user reacts when their views are challenged, and how each user adjusts their perspectives based on the arguments and reasonings that propagate around the network. In other words, when connected together using a Hyperchat AI-enabled platform, the human participants are not just a source of numerical datapoints for aggregation, but are treated as "data processors" who provide reasoning to support their views, who consider the arguments expressed by others, who provide counterarguments to challenge alternate perspectives, and who ultimately converge on a perspective that is likely more informed and thoughtful than they had possessed before the interactive process started.

It is likely that this difference between treating participants as a source of "datapoints" for aggregation versus treating them as "data processors" that engage in thoughtful deliberation is what enables this method to surface a set of forecasts that can significantly outperform traditional Vegas betting markets. In this study, 59 games were conversationally forecast by groups of 20 to 25 participants. The deliberations produced High Confidence forecasts for 27 games, resulting in 78% accuracy versus 57% odds produced by a large-scale Vegas betting market ($p=0.020$). Of the 27 High Confidence predictions, 16 were converged upon through above-average conversation rates. Those 16 picks were 88% accurate versus 57% odds published by large-scale Vegas betting markets, a difference that was statistically significant ($p = 0.010$). These are highly promising results. Future work on Hyperchat AI technology should explore larger conversational groups and determine the impact of scaling-up group size on accuracy.

As we look toward an emerging future of "multi-human, multi-machine" teams, this study illustrates how AI agents can act not merely as passive observers or data aggregators but as active conversational participants that facilitate group deliberation among human teams. As deployed within the Hyperchat architecture, AI agents surface counterpoints, present alternate views and insights, and propagate diverse ideas and reasoning across participants, ensuring that a wide range of perspectives are considered on their merits, while keeping the deliberation productive and balanced. This broadens the use of AI agents in group meetings from being mere transcription tools that summarize discussion to being active collaborators that facilitate ideation, evaluation, assessment, and forecasting. The next frontier lies in orchestrating heterogeneous teams: multiple human groups interacting with multiple AI agents, each with distinct expertise, perspectives, or "personalities." Such architectures will demand new protocols for trust, transparency, and shared agency. Done well, this could unlock new levels of collective intelligence that extend far beyond the sum of its human and machine participants.

With respect to the broader field of collective intelligence and the common methods used to amplify the insights of large groups (i.e., polls, surveys, and markets), this study suggests that large-scale conversational deliberation offers significant advantages over techniques that collect datapoints from individuals and aggregate mathematically. While this study explored the impact of deliberative Hyperchat AI in the context of sports forecasting, it is likely the benefits would extend to other tasks such as business planning, critical decision-making, strategic prioritization, policy evaluation, risk assessment, and deliberative democracy. Future studies should be performed to evaluate these use cases.

And finally, prior collective intelligence research shows that group composition can have a significant impact on collective performance, especially among small deliberative teams [14]. Combining this insight with the capabilities of Hyperchat AI, we can imagine a system that dynamically adjusts the composition of each of the real-time subgroups to maximize performance. Future research should explore this angle.